\documentstyle[psfig, epsfig, 12pt]{article} 
\input{epsf.sty}

\def\mypagenumber{1}

\def\myend{\end{document}}

\normalsize

\newcounter{sxn}

\newcounter{axn}

\date{}

\newdimen\mybaselineskip
\newcommand{\beeq}{\begin{equation}}
\newcommand{\eneq}{\end{equation}}
\newcommand{\be}{\begin{eqnarray}}
\newcommand{\ee}{\end{eqnarray}}
\newcommand{\bpic}{\begin{picture}}
\newcommand{\epic}{\end{picture}}

\def\d{\partial}
\def\dd{\partial}
\def\la{\raise.16ex\hbox{$\langle$} \, }
\def\ra{\, \raise.16ex\hbox{$\rangle$} }

\def\psibar{ \psi \kern-.65em\raise.6em\hbox{$-$} }
\def\mbar{ m \kern-.78em\raise.4em\hbox{$-$}\lower.4em\hbox{} }

\def\n@space{\nulldelimiterspace=0pt \mathsurround=0pt }
\def\huge#1{{\hbox{$\left#1\vbox to 20.5pt{}\right.\n@space$}}}

\def\myskip{\noalign{\kern 8pt}}
\def\myeqspace{\noalign{\kern 10pt}}

\def\boxit#1{$\vcenter{\hrule\hbox{\vrule\kern3pt
    \vbox{\kern3pt\hbox{#1}\kern3pt}\kern3pt\vrule}\hrule}$}
\def\bigbox#1{$\vcenter{\hrule\hbox{\vrule\kern5pt
     \vbox{\kern5pt\hbox{#1}\kern5pt}\kern5pt\vrule}\hrule}$}

\def\ignore#1{{}}

\begin{document}

\bibliographystyle{unsrt}
\footskip 1.0cm

\thispagestyle{empty}
\setcounter{page}{\mypagenumber}

{\baselineskip=10pt \parindent=0pt \small
}                
             
\begin{flushright}{CERN-TH/2001-103}
\\
\end{flushright}

\vskip .8 cm 
\begin{center}
{\LARGE \bf {Deconfinement at $N>2$: $SU(N)$ Georgi-Glashow
model in 2+1 dimensions.}} \\
\vskip 1 cm 
{\large  Ian I.
Kogan$^{a,}$\footnote{kogan@thphys.ox.ac.uk},\hskip 0.2 cm 
 Alex Kovner$^{b,c,}$\footnote{Alexander.Kovner@cern.ch},\hskip 0.2 cm
Bayram Tekin$^{a,}$\footnote{tekin@thphys.ox.ac.uk}
}\\
\vskip 0.3 cm  
$^a${\it Theoretical Physics, University of Oxford, 1 Keble Road, Oxford,
OX1 3NP, UK}\\ 
\vspace{0.5 cm}
$^b${\it Department of Mathematics and Statistics, 
University of Plymouth,
Plymouth PL4 8AA, UK}\\
$^c${\it Theory Division, CERN, CH-1211, Geneva 23, Switzerland}\\

\end{center}

\vspace*{1.5cm}


\begin{abstract}
\baselineskip=18pt  
We analyse the deconfining phase transition in the $SU(N)$ Georgi-Glashow 
model in 2+1 dimensions. We show that the phase transition is second order  
for any $N$, and the universality class is different from the $Z_N$ 
invariant Villain model. At large $N$ the conformal theory 
describing the fixed point 
is a deformed $SU(N)_1$ WZNW model which
has N-1  
massless fields. It is therefore likely that 
its self-dual infrared fixed point is described by the Fateev-Zamolodchikov
theory of $Z_N$ parafermions.

\end{abstract}

\vspace{3.5cm}
CERN-TH/2001-103\\

\vspace{.5cm}

 
\newpage



\normalsize
\baselineskip=22pt plus 1pt minus 1pt
\parindent=25pt

\section{Introduction}
Recently \cite{dunne} we have analysed in detail the deconfining phase 
transition in the $SU(2)$ Georgi-Glashow (GG) model in 2+1 dimensions.
The mechanism of confinement in this model at zero temperature 
is due to the ``plasma'' of the monopole-instantons
and is well
understood \cite{polyakov}. The model is weakly interacting
all the way up to the deconfining temperature, which allowed us
to study the phase transition quantitatively. 
We found that taking into account the excitations of the heavy charged
particles was crucial for the correct description of the transition.
The 
transition is associated with the restoration of the magnetic $Z_2$ symmetry
\cite{thooft1, kovner} in accordance with general arguments of \cite{KOVNERZN}.
The universality class of the transition was found to be 2d Ising.

Whereas for $SU(2)$ gauge theory there is overwhelming consensus that 
the transition should be in the universality class of the Ising model, 
the situation is much less clear for large $N$. The point is that
for $N>3$ one can write down different 2d spin models, and they have
different critical behaviour. For example the $N$-state Potts
models have first order phase transition for $N>4$ \cite{potts}, 
while Villain models
have second order transition which is of the BKT type, and is thus
in the universality class of $U(1)$ \cite{villain}.
Whether the transition in the $SU(N)$ gauge theory is similar to either one
of those, is an open interesting question.

In this paper we consider a general Georgi-Glashow type $SU(N)$ gauge theory,
where at zero temperature the gauge group is spontaneously broken to
$U^{N-1}(1)$. Just like the $SU(2)$ GG model, the theory is weakly interacting.
At zero temperature it is confining, and the monopole ``plasma''
description of confinement has long been known \cite{wadia}.
It has also been studied from the point of view of magnetic $Z_N$ symmetry 
in \cite{snyderman}.

Our main finding is that the transition in the model is second order, and
is distinct from that of Villain model. Although we are unable to identify
the fixed point theory with a known two dimensional conformal theory, we
argue that the relevant model at large $N$ must be a deformation of a theory 
with  a large value of the  UV central charge $c=O(N)$, which may be $SU(N)_1$ WZNW model.

The paper is structured as follows. In Section 2 we describe the model 
as well as the monopole and magnetic symmetry based approaches to its
low energy dynamics. In Section 3 we derive the dimensionally reduced
model relevant for the study of the phase transition, and discuss the
role of the heavy charged particles. In Section 4 we study the transition
with the help of the renormalization group analysis in the reduced theory.
We show that the RG equations have a self dual infrared fixed point.
We explain why the GG model close to the transition
does not behave like Villain model, even in the
range of parameters where one might expect it to do so.
In Section 5 we point out to similarities between the behaviour of some
quantities in the GG model close to criticality and in the hot 
Yang Mills theory.  Finally in Section 6 we discuss our results.

\section{The model}
We consider the $SU(N)$ gauge theory with  scalar fields in the
adjoint representation in 2+1 dimensions. 
\be
{\cal{L}}= -{1\over 2} \mbox{tr}F_{\mu \nu}F^{\mu \nu} +
 \mbox{tr}D_\mu \Phi D^\mu \Phi  -  V(\Phi)
\label{model1}
\ee
where
\be
&&A_\mu = A^a_\mu T^a \hskip 1 cm 
F_{\mu\nu} = \dd_\mu A_\nu  -\dd_\nu A_\mu +  g[A_\mu, A_\nu] \nonumber \\ 
&&\Phi = \Phi^a T^a \hskip 1 cm  D_\mu \Phi = \dd_\mu \Phi + g [A_\mu, \Phi]. 
\ee
$T^a$ are
traceless hermitian generators of the $SU(N)$ algebra 
normalised as $\mbox{tr}(T^a T^b)={1\over 2}\delta^{ab}$. 

Depending on the form of the Higgs potential, there 
can be different patterns of gauge symmetry breaking. 
Since most of the details of the potential are unimportant for our purposes,
we will not specify it except for restricting it to the region of the
parameter space where classically the
gauge symmetry is broken to the maximal torus
\be
SU(N) \rightarrow U(1)^{N-1}
\label{symmetry}
\ee
We also restrict ourselves to weakly coupled regime, which means that the
ratios $M_W/g^2$ are large for {\it all} $N^2-N$ massive $W$-bosons.

\subsection{The perturbative spectrum.}
To characterise the perturbative spectrum of the theory it is
convenient to use
the Cartan-Weyl basis $(H^i, E^{\vec{\alpha}})$, where $H^i$ generate the
Cartan subalgebra which is of the dimension of rank of $SU(N)$: $r= N-1$.
\be
[H^i,H^j]=0  \hskip 1 cm i,j \in [1,2,.. N-1]
\ee
and $E^{\vec{\alpha}}$ are the $N(N-1)$ ladder operators 
which satisfy
\be
&&[H^i, E^{\vec{\alpha}}]= \alpha^i E^{\vec{\alpha}}, \\ 
&& [E^{\vec{\alpha}},E^{\vec{\beta}}]= N_{\vec{\alpha},\vec{\beta}}\,\, E^{\vec{\alpha}+\vec{\beta}} 
\hskip 0.5 cm \mbox{if}\,\,\, \vec{\alpha}+\vec{\beta} \,\,\,\mbox{is a root} \\  
&& \hskip 1.5 cm =  2\vec{\alpha} \cdot \vec{H} \hskip 0.5 cm \mbox{if} \,\,\,\,\vec{\alpha} = - \vec{\beta}
\ee
The $N-1$ dimensional 
root vectors  $\vec{\alpha} = (\alpha^1,\alpha^2,... \alpha^{N-1})$ 
form the dual Cartan subalgebra. There are obviously $N(N-1)$ 
such vectors corresponding to 
$\mbox{dim}(SU(N))- \mbox{rank}(SU(N))$ but only $N-1$ of them are linearly 
independent.  
The non-vanishing inner products in the Cartan-Weyl basis read as
\be
\mbox{tr}(H^i,H^j) = {1\over 2}\delta^{i j}, 
\hskip 1 cm  \mbox{tr}(E^{\vec{\alpha}},E^{\vec{\beta}}) = 
{1\over 2}\delta^{\vec{\alpha},-\vec{\beta}}.
\ee

At the classical level $N-1$ gauge group generators 
are unbroken, which we choose to correspond to $(H^i)$. Therefore classically 
there are $N-1$ massless photons and $N(N-1)$ charged massive W-bosons. 

Our Weyl basis is chosen in such a way that the Higgs VEV is diagonal. Since
the matrix $\Phi$ is traceless, there are $N-1$ independent eigenvalues.
In terms of the $N-1$ dimensional vector 
$\vec{h}= (h_1, h_2, h_3,..h_{N-1})$\footnote{
For concreteness we order these numbers $h_1 \,\,> \,\, h_2 > ... h_{N-2} \,\,>\,\,  h_{N-1}$, which also breaks the discrete Weyl group.}
we have
\be
<\Phi> =  \vec{h} \cdot \vec{H} , \ \ \ \ \ \ A_\mu =    \vec{A}_\mu \cdot \vec{H} + \sum_{\vec{\vec{\alpha}}} A^{\vec{\alpha}}_\mu E^{\vec{\alpha}}
\ee
For concreteness let us choose the following basis for the Cartan subalgebra;
\be
&&H_1= {1\over 2}\mbox{diag}(1,-1, 0, ...0), \hskip 0.5 cm H_2= {1\over 2\sqrt{3}}\mbox{diag}(1,1, -2, 0 ...0) \nonumber \\
&& \,\,\,\, ... \ \ \ H_{N-1}= {1\over\sqrt{2N(N-1)}}\mbox{diag}(1,1, 1, ...1, - (N-1))
\ee
As long as $\vec{h} \cdot \vec{\alpha} \neq 0$ for all roots, 
the gauge symmetry is maximally broken.
The masses of the W-bosons can be read off from the second term in the lagrangian
\be
&& g^2 \mbox{tr}[A_\mu,\Phi]^2 = {g^2\over 2}\sum_{\vec{\alpha},i,j} 
A^{\vec{\alpha}}_{\mu}A^{-\vec{\alpha}}_{\mu}h_i h_j \alpha^i \alpha^j \\
&&\Longrightarrow  M_{\vec{\alpha}} = g |\vec{h} \cdot \vec{\alpha}| 
\ee

The W-bosons corresponding to the $N-1$ simple roots $\vec\beta_i, \ \ \ 
i=1,...,N-1$
(arbitrarily chosen set of linearly independent roots)
can be thought of as fundamental,
in the sense that the quantum numbers and 
the masses of all other W-bosons are obtained as linear combinations
of
those of the fundamental W-bosons. These charges and masses are
\be
\vec{Q}_{\vec{\beta}}= g\vec{\beta}, \hskip 1 cm M_{\vec{\beta}} = g \vec{h} \cdot \vec{\beta} 
\label{cham}
\ee
As an example consider the case of $SU(3)$ broken down to $U(1)\times U(1)$.
There are 6 massive W-bosons. 
The simple roots can be taken as
\be
\vec{\beta}_1 =  ({1\over 2}, {\sqrt{3}\over 2}), \hskip 1cm \vec{\beta}_2 =  
(-{1\over 2}, {\sqrt{3}\over 2})
\label{simpleroots}
\ee
The remaining non-simple positive root is
\be
\vec{\alpha}_3 = \vec{\beta}_1 - \vec{\beta}_2 = (1,0).
\ee
The other three roots are $-\vec\beta_i$, $-\vec\alpha_3$.
The masses of corresponding W-bosons are
\be
M_{W_1}= {g\over 2}(h_1 +\sqrt{3} h_2 ), \hskip 0.5 cm M_{W_2}= {g\over 2}(h_1 -\sqrt{3} h_2 ), \hskip 0.5 cm
M_{W_2}= gh_1
\label{wmass}
\ee
for $h_1 > \sqrt{3} h_2$.
Observe that if $h_2 = 0$, two of the masses become degenerate. 
In this case $SU(3)$ is still broken down to $U(1)\times U(1)$
since all three masses are non-vanishing but the
spectrum is invariant under an additional  $Z_2$ symmetry.
This $Z_2$ symmetry is the charge conjugation with respect to the charge $H_2$, which interchanges the roots $\vec\beta_1$ and $\vec\beta_2$.
In general though, this charge conjugation symmetry is broken
by the VEV of Higgs \cite{chris}.

\subsection{The monopole-instantons and the Polyakov effective Lagrangian.}
Non-perturbatively the most important contributions in the theory are due to the
monopole-instantons.
Those are classical, stable, finite action solutions of the Euclidean equations of motion
arising due to the nontrivial nature of the 
second homotopy group of the vacuum manifold 
($\Pi_2(SU(N)/U(1)^{N-1}) = Z^N$).
The magnetic field of such a monopole is long range.
\be
B_\mu = {x^\mu \over 4 \pi r^3} \vec{g} \cdot \vec{H} 
\ee
The $N-1$ dimensional vectors $\vec{g}$ are determined by
the non-Abelian generalisation of the Dirac quantisation condition 
 \cite{englert, goddard}
\be
e^{i g\vec{g}\cdot \vec{H} }= I 
\ee
Solution of this quantisation condition are
\be
\vec{g}= {4\pi \over g} \sum_{i= 1}^{N-1} n_i \vec{\beta}^*_i  
\ee
where $\vec{\beta}^*$ are the 
dual roots defined by $\vec{\beta}^* = \vec{\beta}/ |\vec{\beta}|^2$. 
We will be working with roots normalised to unity, and thus 
$\vec{\beta}^* = \vec{\beta}$.
The integers $n_i$ are elements of the group  
$\Pi_2$ \cite{weinberg}. The monopoles which have the smallest action
correspond to roots taken once. The action of these
 monopoles in the BPS limit is
\be
M_{\vec\alpha} = {4 \pi \over g}\vec{h} \cdot \vec{\alpha}=
{4\pi M_{W_{\vec\alpha}}\over g^2}
\ee
Just like with $W$-bosons we can think of monopoles corresponding to simple roots as
fundamental ones with magnetic charges and action
\be
 \vec{g_i}= {4\pi \over g}\vec{\beta}_i  \hskip 1 cm   M_i = {4 \pi \over g}\vec{h} \cdot \vec{\beta}_i   
\ee
 For example, in the case of $SU(3)$ (see eqs.(\ref{simpleroots},\ref{wmass})) 
the monopole action  spectrum (in the BPS limit) is 
\be
M_1= {2\pi \over g}(h_1 +\sqrt{3} h_2 ), \hskip 0.5 cm M_2= {2\pi\over g}(h_1 -\sqrt{3} h_2 ), \hskip 0.5 cm
M_3= {4\pi\over g}h_1.
\ee

The effect of these monopoles is to impart finite mass to all the 
perturbatively massless ``photons''. The derivation of the effective 
Lagrangian follows exactly the same lines as the original derivation 
of Polyakov for the $SU(2)$ theory \cite{polyakov}. The resulting low
energy effective theory is
written in terms of the $N-1$ component field, $\vec{\eta}$, 
with the
following Lagrangian \cite{wadia, snyderman} 
\be
{\cal{L}}_{eff} = {g^2\over 32\pi^2} (\d_{\mu}\vec{\eta})^2 + 
\sum_\alpha {M_\alpha^2 g^2\over 16\pi^2 }\mbox{exp}(i\vec{\alpha} \cdot \vec{\eta})
\label{lowenergy}
\ee
The sum is over all $N(N-1)$ non vanishing roots. 
The potential induced by the monopoles is proportional to the 
monopole fugacity
\be
M^2_\alpha= {16 \pi^2 \xi_{\alpha}\over{g^2}}, \hskip 2 cm \xi_{\alpha} = \mbox{constant}
{M_{W_\alpha}^{7/2}\over g} e^{-{4\pi M_{W_\alpha} \over g^2} 
\epsilon({M_H\over M_W})}.    
\ee
$\epsilon({M_H\over M_W})$ is such that  $ 1\leq \epsilon \leq
1.787$ \cite{prasad}, and $\epsilon(\infty)=1$.

The photons at weak coupling are obviously much lighter than the $W$ - bosons
and thus are the only relevant degrees of freedom in the low energy sector.

\subsection{The magnetic $Z_N$ symmetry.}

The global symmetry structure is very important for the understanding of the
deconfining transition. The relevant symmetry in the present model is the
magnetic $Z_N$ symmetry. We now wish to explain how this symmetry is
implemented in the effective low energy Lagrangian.
Our discussion parallels the $SU(2)$ case \cite{dunne}.

The order parameter of the magnetic symmetry is the set of magnetic vortex 
operators $V_i$, $i=1,...,N-1$. These operators were constructed explicitly in
\cite{kovner}. These operators carry magnetic fluxes of the $N-1$ $U(1)$ 
Abelian magnetic fields. The defining commutation relation 
for $V_i$ is
\be
[V_i(x),{\vec B}(y)]=-{4\pi\over g}{\vec w}_iV_i(x)\delta^2(x-y)
\label{nvortex}
\ee
Here ${\vec B}$ is the $N-1$ dimensional vector of magnetic 
fields\footnote{Note that these magnetic fields can be
constructed in gauge invariant way from the non-Abelian field strengths and
the Higgs field, see \cite{kovner}.}, whose
$j$-th component is the projection of the non-Abelian field strength onto 
the direction of the Cartan subalgebra generator $H_j$, and ${\vec w}_j$ are 
$N-1$ weight vectors of $SU(N)$. The choice of the $N-1$ out of $N$ weight 
vectors
is arbitrary. Change in this choice will lead to the redefinition of the 
vortex operators such that the new operators will be products of the old ones
and their conjugates.
It is always possible to choose these weights so that together with the
``fundamental'' roots $\beta_i$ they satisfy the relation
\be
{\vec w_i}{\vec \beta_j}={1\over 2}\delta_{ij}
\label{relat}
\ee
The flux eigenvalues in eq.(\ref{nvortex}) are dictated by the requirement of
the locality of the vortex operators and is analogous to the Dirac 
quantisation condition. The explicit form of the vortex operators
in terms of the field $\eta$ in eq.(\ref{lowenergy}) is
\be
V_i(\vec{x})= {g\over\sqrt{8} \pi}\,e^{i\chi_i}
\ee
with
\be
\chi_i =  \vec{w}_i \cdot \vec{\eta}  \Longrightarrow \vec{\eta} = 2 \sum_i \vec{\beta}_i \chi_i
\label{chieta}
\ee
The effective Lagrangian can be written as a nonlinear $\sigma$-model in terms of $V_i$ as
\be
{\cal{L}}_{eff} = {N-1\over 2}\sum_{i,j} A_{ij} {1\over V_k^*V_k}(V_i^*\d_{\mu}V_i)(V_j \d_{\mu}V_j^*) + 
+\lambda(\sum_i (V_i V_i^* - {g^2\over 8\pi^2})^2 + \sum_{\alpha} k_{\alpha}\,\prod_i V_i^{2 i\vec{\alpha} \cdot \vec{\beta}_i}  
\label{vlag}
\ee
with $\lambda\rightarrow\infty$.
The matrix $A_{ij}= 2\vec{\beta}^*_i \cdot \vec{\beta}_j$ 
depends on the choice of the fundamental roots. With the conventional 
choice of positive roots, where $\vec\beta_i\vec\beta_j=-1/2, \ \ \ i\ne j$,
it
is the 
Cartan matrix of the Lie algebra. All its diagonal elements are equal
to $2$, while all its off diagonal elements equal to $-1$.
We will find it however more convenient in the following to use a different 
set of fundamental roots, for which  
$\vec\beta_i\vec\beta_j=1/2, \ \ \ i\ne j$. Such a choice is always possible
for any $SU(N)$. For this choice of roots the off-diagonal matrix elements
of $A_{ij}$ are all equal to $1$.

For $SU(3)$ we have 
\be
{\bf A}=\left ( \begin{array}{cc}
2 & 1\\
1& 2 
\end{array}
\right ) 
\ee
and the effective Lagrangian
\be
{\cal{L}}_{eff}=  &&\d_{\mu}V_1 \d_{\mu}V_1^* + {8\pi^2\over g^2}V_1^*\d_{\mu}V_1 V_2 \d_{\mu}V_2^* +  \d_{\mu}V_2 \d_{\mu}V_2^* 
\nonumber \\
&&+ \xi_1( V_1 V^*_2 + \mbox{c.c} ) +  \xi_2( V_1^2 V_2 + \mbox{c.c} )
+\xi_3( V_1 V_2^2 + \mbox{c.c} )
\label{vortexlag}
\ee

The magnetic $Z_N$ symmetry has an obvious and simple representation in this
effective Lagrangian as $V_i\rightarrow\exp\{2\pi in/N\}V_i$.

As long as only small fluctuations of the phase fields $\chi_i$ are
important, the Lagrangian eq.(\ref{vortexlag}) is equivalent to the
eq.(\ref{lowenergy}). Thus at low temperature the descriptions based on these
Lagrangians are equivalent. The difference appears only when the phase 
nature of $\chi_i$ plays a role. Indeed, since $\chi_i$ are treated 
in eq.(\ref{vortexlag}) as phases, dynamically one allows configurations in
which these phases have nontrivial winding. On the other hand 
in eq.(\ref{lowenergy}) such configurations cost infinite amount
of energy. As discussed in detail in \cite{kovner} and \cite{dunne} the
winding configurations correspond to the heavy $W$-bosons. 
In fact the explicit relation between the vorticity of the fields $V_i$ and the
electric charges is given by \cite{kovner}
\be
{1\over g}{\vec w}_i{\vec Q}={1\over 4\pi}\oint_{C\rightarrow\infty}dl_\mu
\partial_\mu\chi^i_\mu
\label{vorticity}
\ee
Thus the
difference between the two Lagrangians is important whenever
the physics of the $W$ bosons plays an important role. We have seen in the
case of the $SU(2)$ theory that $W$'s are indeed important near the phase
transition temperature. The same turns out to be true for arbitrary
$SU(N)$. We thus have to be careful to treat the $W$ - bosons 
properly in the transition region. In the next section we will set up
this treatment.

\section{The reduced theory.}
Throughout this paper we are working in the weak coupling regime and
thus the photon masses in eq.(\ref{lowenergy}) are exponentially small. Thus
already at very low temperature ($T\propto M_\alpha$) one can use the 
dimensionally
reduced version of the theory, since all the thermal modes are significantly
heavier than the zero Matsubara frequency mode. Since the critical
temperature for the deconfining transition is of order $g^2$ 
(see \cite{dunne}), we can safely use dimensional reduction close to the
transition. The zero Matsubara frequency sector is described by
the {\it two} dimensional Lagrangian
\be
{\cal{L}}_{eff} = {g^2\over 32\pi^2 T} (\d_{\mu}\vec{\eta})^2 + 
\sum_\alpha {M_\alpha^2 g^2\over 16\pi^2 T }\mbox{exp}(i\vec{\alpha} \cdot \vec{\eta})
\label{reduced}
\ee
However, as we noted before, our description should include $W$ bosons, and so
the fields $\eta$ should be treated here as phases with periodicity appropriate
to eq.(\ref{chieta}). In fact the Lagrangian also has to be augmented by
a four derivative ``Skyrme'' term, which fixes the energy of the winding states
to be equal to the masses of $W$ bosons \cite{dunne}.
We can however simplify things further, by noting that the density of
$W$ bosons at criticality is exponentially small due to the Boltzmann factor
suppression. Thus $W$'s can be treated in the dilute gas approximation
in the same way as was done in \cite{dunne}.
To do this explicitly we first have to understand how to write partition
function in the presence of one $W$ boson of a particular type.

Let us first consider a $W$ boson corresponding to one of the fundamental
roots $\beta_k$.
Using eq.(\ref{cham}), eq.(\ref{relat}) and eq.(\ref{vorticity}) 
we see that this $W$ boson corresponds to unit vorticity of the field $V_k$
and zero vorticity of all other fields $V_j$, $j\ne k$.
To create such a vortex in the path integral we must introduce an external
``current''which forces the discontinuity
of the field $\chi_k$
\be
\chi_k = \chi_k + 2\pi
\ee
The partition function in the presence of one $W$ boson is thus
\be
&&Z = \int D[\chi(x)] \nonumber \\
&&\exp\left\{-\int d^2 y { g^2 \over 16 \pi^2 T}\sum A_{ij}
(\d_{\mu}\chi_i-
J^i_\mu(y,x))(\d_{\mu}\chi_j-
J^j_\mu(y,x)) + \sum_{\alpha}\zeta_{\alpha} \cos ( i 2 \sum_i \vec{\alpha}\cdot \vec{\beta}_i \chi_i)
\right \}
\nonumber
\label{onev}
\ee
with
\be
J^i_{\mu}(y,x)=2\pi\delta_{ik} n_\mu(y)\delta(y\in C_x)
\ee
with $C_x$ a curve that starts at the location of the vortex (the point
$x$), and goes
to infinity, and $n_\mu$ is the unit normal to this curve.
The insertion of this current forces the normal derivative of $\chi_k$ to 
diverge on curve $C$, so that $\chi_k$ jumps
by $2\pi$ across $C$. Since in the rest of the space $\chi_k$ is
smooth, the path integral is dominated by a configuration with unit vorticity of $\chi_k$\footnote{Note that even though $J^i_\mu$ explicitly depends on the curve $C_x$, the
partition function itself does not, since changing the integration
variable $
\chi_i(x)\rightarrow\chi_i(x)+2\pi , \ \ \ \ \ \ \ x\in S$
where the boundary of $S$ is $C_x-C_x'$ is equivalent to changing $C-x$ into
$C_x'$ in the definition of the current.}.

The path integral eq.(\ref{onev}) differs from the partition function
in the vacuum sector by the linear term in the Lagrangian
\be
- { g^2 \over 4 \pi^2 T } \sum_{i,j}\int d^2 y\, \vec{\beta}_i\cdot \vec{\beta}_j\, \partial_{\mu}\chi_i J^j_{\mu} 
= - { g^2 \over 4 \pi T }
\int_{C_x} dx_\mu\epsilon_{\mu\nu}\,\partial_\nu\vec{\beta_k} \cdot \vec\eta
\ee
Defining in the standard way the dual field $\tilde\eta$,
\be
i\d_{\mu}\vec{\tilde{\eta}} = \epsilon_{\mu \nu} \d_{\nu}\vec{\eta}
\ee
we can recast the contribution of this particular $W$ boson
in the form of the following extra term in the Lagrangian
\be
- i{ g^2 \over 4 \pi T }\vec{\beta_k}\cdot \vec{\tilde{\eta}}
\label{wcont}
\ee
This procedure can be repeated for $W$ boson corresponding to an arbitrary
root $\alpha$ with the only difference that in eq.(\ref{wcont}), the root
$\beta_k$ is replaced by the root $\alpha$.
To create several $W$-bosons one just inserts the external current which
is the sum of the currents creating individual $W$'s. 

Dilute ensemble of such objects with small fugacities $\mu_{\alpha}$
is then given by
\be
Z= \prod_{\alpha} \sum_{n,m}{1\over n!}{1\over m!}\mu_{\alpha}^{n+m}\int \prod_idx_i\prod_jdy_j
Z(x_i,y_j)
\ee
The summation over the number of $W$'s can be
easily performed, see \cite{dunne}. The result is 
the partition function with the Lagrangian
\be
{\cal{L}}_{eff} = {g^2\over 32\pi^2 T} (\d_{\mu}\vec{\eta})^2 + 
\sum_\alpha \zeta_{\alpha}\mbox{exp}(i \vec{\alpha} \cdot \vec{\eta})
+\sum_\alpha \mu_{\alpha}\mbox{exp}(i {g^2\over 4\pi T} \vec{\alpha} \cdot \vec{\tilde{\eta}})
\label{lowenergy2}
\ee
with summation in both terms going over all non-vanishing roots of $SU(N)$.
The coefficients $\mu_\alpha$ are proportional to the fugacities of the
corresponding $W$ bosons
\be
\mu_\alpha\propto \exp\{-M_{W_\alpha}/T\}
\ee
Eq.(\ref{lowenergy2}) is the dimensionally reduced theory which we will now use
to study the phase transition.

\section{The phase transition.}
\subsection{Monopoles versus charges.}
To study the phase transition we may first attempt to disregard the 
$W$ boson induced term in the effective Lagrangian. If we do that,
 we are back to
the theory eq.({\ref{reduced}). This theory is easily analysed.
The first interesting thing about it is that 
since the group is simply laced 
(all the roots are of unit length) the anomalous dimensions of all the 
interaction terms are equal.
The scaling dimension of all the monopole induced terms is
\be
\Delta_M = {4\pi T\over g^2}
\ee
This immediately tells us that at the temperature
\be
T_{BKT}= {g^2\over 2\pi}
\ee
all these interactions become irrelevant. Thus at $T_{BKT}$ one
 expects the Berezinsky-Kosterlitz-Thouless transition to take place.
Above this temperature the infrared behaviour of the theory is that of 
$N-1$ free massless particles. Note that $T_{BKT}$ does not depend 
on the number of colours $N$. If the picture just described where true,
the universality class of the phase transition would be
that of $U^{N-1}(1)$. 

This of course is exactly the same situation as encountered in \cite{zarembo}
in the $SU(2)$ case. Again just like in $SU(2)$ case this conclusion is
incorrect due to the contribution of the $W$ bosons.
To see this it is simplest to ask what would happen at high temperature
if there were no monopole contributions at all. This amounts to studying
eq.(\ref{lowenergy2}) with $\xi_\alpha=0$. This theory describes non-compact
electrodynamics with $N-1$ photons and the spectrum of charged particles 
given by eq.(\ref{cham}). This limit is again simple to understand, 
since the theory is exactly dual to the theory with monopoles and without 
charges. The scaling dimensions of all the $W$ induced perturbations
are equal and are given by
\be
\Delta_W = {g^2\over 4\pi T}
\ee
Thus the perturbations are irrelevant at low temperature, but become
relevant at 
\be
T_{NC}={g^2\over 8\pi}
\ee
Since $T_{NC}<T_{BKT}$ this tells us that we can not neglect the effects 
of charges at criticality.
The story of $SU(2)$ exactly repeats itself. Even the
value of the temperature at which
the scaling dimensions of the charge- and monopole induced perturbations
are equal does not depend on $N$.

We expect therefore that the actual transition temperature is
\be
T_C={g^2\over 4\pi}
\ee
at which point all perturbations have the same scaling dimension.
This expectation is confirmed by the renormalization group analysis.

\subsection{Renormalization group analysis}
The renormalization group equations for the theory eq.(\ref{lowenergy2})
were studied in \cite{boyanovsky}. 
In general the equations are quite complicated 
due to the cross correlations between different operators. For this reason
 the space of
parameters of the theory has to be enlarged if one wants to study 
the flow whose UV initial condition is provided by eq.(\ref{lowenergy2})
with arbitrary values of fugacities. However there is one simple
case, that is when the initial condition is such that all the monopole 
fugacities are equal $\xi_{\alpha_i}=\xi_{\alpha_j}=\xi$, 
and all the charge fugacities are equal $\mu_{\alpha_i}=\mu_{\alpha_j}=\mu$.
This initial condition is stable under the RG flow. On this subspace 
the RG equations, written in terms of the scaled temperature
$t={4\pi T\over g^2}$ and dimensionless fugacities, read
\be
&&{\d t\over \d \lambda} = 2 \pi^2 N t( \mu^2 - \zeta^2 ) \\
&& {\d \mu \over \d \lambda} = (2 - {1\over t}) \mu  - 2 \pi(N-2) \mu^2 \\
&&{\d \zeta \over \d \lambda} = (2 -  t) \zeta  - 2 \pi(N-2) \zeta^2 
\label{rge}
\ee
These equations have exactly the property reflecting our previous discussion.
That is the points $t=2, \ \ \mu=0$ and $t=1/2, \ \ \ \xi=0$ are both unstable.
The stable IR fixed point is
\be
t_o = 1\hskip 1 cm \mu_0= \zeta_0 = {1 \over 2\pi(N-2)}  
\label{fixedp}
\ee
One can in fact easily check that in the three dimensional space of couplings
$t,\xi$ and $\mu$ this point has two attractive and one repulsive direction.
This is precisely what one expects from the IR fixed point located on the
critical surface, the two attractive directions being the tangential directions
to the surface.

The RG equations have an obvious duality symmetry, $\mu\rightarrow\xi, \ \ \ 
t\rightarrow 1/t$. This is the reflection of the transformation 
$\eta\rightarrow\tilde\eta$ on the level of 
the Lagrangian eq.(\ref{lowenergy2}).
The points $t=1$, $\mu=\xi$ are symmetric under duality, and this ensures
existence of a self dual fixed point. This is important, since
the 
exact position of the fixed point is scheme dependent. Its existence however
is assured by the duality symmetry. 

What is the nature of this fixed point? For $N=2$ we were able in \cite{dunne}
to fermionize the fixed point theory and show explicitly that it is
equivalent to one massless Majorana fermion. We are not able to perform
a similar analysis for arbitrary $N$.
There are however several comments that we would like to make.
Phase transitions in $Z_N$ invariant spin models have been studied quite
extensively. 
A nice recent discussion of the situation is given in \cite{znspin}. 
One considers a spin model of one phase field $\theta$ with a symmetry breaking
term of the type 
$h\ \cos\{N\theta\}$ which breaks the $U(1)$ symmetry down to $Z_N$. When the
coefficient $h$ of this symmetry breaking term is large, the model resembles
Potts model and thus (for $N>4$) has a first order phase transition. When the 
breaking is small on the other hand, the behaviour is similar to the Villain 
model: the system undergoes two BKT type transitions with a massless
$U(1)$ symmetric phase at intermediate temperatures. At some particular
``tricritical'' 
value of $h$ the massless phase shrinks to a point and it comes together 
with the first order transition line. This tricritical point is self-dual and
is described by a conformal $Z_N$ invariant parafermionic 
theory with the central charge
$c=2(N-1)/(N+2)$ introduced in \cite{zamolodchikov}.
In this type of models  
therefore generically one expects either the first order 
transition or a pair of BKT transitions with the massless phase in between.
The tricritical behaviour is special and requires fine tuning
of the parameters.
This is indeed also the prevailing general expectation for the order of 
the transition in 2+1 dimensional gauge theories at large $N$: either first
order or Villain type $U(1)$ invariant behaviour.

In fact we find our model in a completely different situation.
The transition is not first order, and there is no $U(1)$ invariant
massless phase. We stress that within the RG flow eq.(\ref{rge}) the 
IR fixed point eq.(\ref{fixedp}) has two attractive directions. This
means that it governs the IR behaviour of the points which lay
on 2 - dimensional
critical surface in the three dimensional parameter space, and is 
therefore generic.
This by itself does not preclude that this fixed point is the same as the
parafermionic $Z_N$ theory of \cite{zamolodchikov}. If this is the case,
it is quite interesting, since the point which appeared as ``tricritical'' from
the point of view of usual spin models is in fact generic
from the point of view of the 3D gauge theories.  At present we can not prove 
 that our critical point is 
described by the parafermionic theory  but let us present some arguments 
 supporting this conjecture.
 The point is that, as
opposed to models considered in \cite{znspin} 
our Lagrangian eq.(\ref{lowenergy2})
describes a theory of $N-1$ light fields. 
The theory of $N-1$
free massless fields have the UV central charge $c_{UV} = N-1$.
However this CFT is deformed by the monopole and $W$ - induced perturbations
and flows to a different IR fixed point.
However let us note that the central charge $c=N-1$
is precisely the central charge of the $SU(N)_1$ WZNW model. 
The Ising 
(i.e. $c=1/2$) model  is the lowest among the 
minimal models with Virasoro (i.e.
$W_2$)  symmetry. The highest model of this class 
is $c=1$ model (one free field ) which is  precisely
 $SU(2)_1$ WZNW model.
When the $c=1$ model was deformed by the monopole 
and $W$-boson  operators the central charge was reduced -
and the resulting IR theory was  Ising.

Now,  $Z_N$ parafermions with $c = 2(N-1)/N+2$  are the lowest minimal 
 models with $W_N$ symmetry  - and the highest is  
$SU(N)_1$  (for more information about parafermions  
see for example  \cite{cappelli} and references therein) 
which can be described  in terms  of 
 $N-1$  massless  fields.
Thus if the theory in the UV describes N-1  massless fields  and has $W_N$
symmetry, it is quite possible  that result of 
the relevant (monopole+$W$)  deformation is a self-dual critical
point. It is indeed known 
that the $Z_N$ parafermion theory is the self-dual model with $W_N$
symmetry. The fact that the central charge (and thus the effective 
number of degrees of freedom) is reduced in the process of the flow towards IR
is of course in complete accord with Zamolodchikov's C-theorem.
It is therefore possible that the IR fixed point that describes 
the universality
class of the GG model is the conformal $Z_N$ parafermion theory.

Analysis of \cite{boyanovsky}, 
although
admittedly incomplete also supports the expectation that we do not 
have Villain picture.
In fact it is the presence of the large number of fields that drives our theory
away from the Villain behaviour as we will now explain.

\subsection{Why not Villain?}

The RG equations eq.(\ref{rge}) were derived for the situation where
all $W$ bosons have equal masses 
(all fugacities $\mu_\alpha$ are equal). One can
wonder what happens if this is not so. In particular imagine an extreme 
situation, where some $W$ bosons are light relative to the others, so that
large monopole fugacities make all phase fields 
$\chi_i$ (or components of $\vec\eta$)
but one relatively heavy.
In this case at zero temperature the theory 
seems to have only one light degree of freedom. This situation is
as close as it can be to the spin systems with one phase field, and one
may expect that in this region of parameter space the finite temperature 
behaviour will be similar to that in the Villain model.
The appearance of the intermediate massless phase potentially has a natural
place in our model. It could occur if the temperature at which the monopoles
become irrelevant 
is lower than the temperature at which charges become relevant,
$T_{BKT}<T_{NC}$. Then between these 
two temperatures the theory in the infrared
is the theory of massless photons. Indeed, consider a simple
$Z_N$ invariant theory of one phase field
\be
{\cal{L}}_{eff} = {g^2\over 8\pi^2 T} (\d_{\mu}\phi)^2 + 
\xi\mbox{exp}(iN\phi)
\ee
We normalised the kinetic term so that for $N=2$ the model reduces to the
Polyakov effective theory for $SU(2)$ GG theory.
The BKT point in this theory is at
\be
T_{BKT}={2g^2\over \pi N^2}
\label{naivebkt}
\ee
If the only vortices that are allowed have integer vorticity, the
temperature at which they become relevant does not depend on $N$ and is
\be
T_{NC}={g^2\over 8\pi}
\label{naivenc}
\ee
Thus for $N>4$ the ``monopole binding'' occurs prior to the
``charge deconfinement'' and there is an intermediate massless phase,
bounded by two BKT transitions.

Let us analyse in more detail how the model eq.(\ref{lowenergy2})
behaves when one
photon is much lighter than the others. The simplest case is $SU(3)$
eq.(\ref{vortexlag}). Let us take $W_1$ to be lighter than $W_2$ and $W_3$.
This means that in eq.(\ref{vortexlag}) we have $\xi_1\gg\xi_2,\ \ \xi_3$.
To minimise the first term in the potential, dynamically the difference 
of the phases of the two vortex fields must be constant. 
Thus on the low energy states we have
\be
V_1=V_2
\ee
With this identification we indeed get the theory of one phase field. 
However the coefficient of the kinetic term is ``renormalised'' due to the
off diagonal form of $A_{ij}$. In this case we find
\be
{\cal{L}}_{eff} = {3g^2\over 8\pi^2 T} (\d_{\mu}\chi)^2 + 
\xi\mbox{exp}(i3\chi)
\ee
This reduction procedure is easily extended to any $N$. One can always
choose appropriate $W$''s to be light, so that at low
energy {\it all} vortex fields become equal
\be
V_i=V_j
\label{equal}
\ee
The effective theory then is
\be
{\cal{L}}_{eff} = {N(N-1)g^2\over 16\pi^2 T} (\d_{\mu}\chi)^2 + 
\xi\mbox{exp}(iN\chi)
\label{lred}
\ee
Interestingly, the coefficient of the kinetic term of the only remaining
field is of order $N^2$, which is the number of degrees of freedom in the 
underlying Yang-Mills theory.
Thus the first thing to note is that the BKT temperature does not
decrease as suggested by eq.(\ref{naivebkt}), but rather increases with $N$ as
\be
T_{BKT}={g^2(N-1)\over \pi N}
\ee
so that at $N\rightarrow\infty$ its value is twice that of $N=2$.

To calculate $T_{NC}$ we should look at the terms that contain dual 
fields in eq.({\ref{lowenergy2}). The structure 
of the phases in these terms is exactly the same as the structure
of the phases in the monopole induced term. Thus clearly taking all $\chi_i$ 
(and therefore $\tilde\chi_i$) to be equal some of these phases will vanish, 
while others will give the only surviving $\tilde\chi$ field with the 
coefficient $N$.
Thus the charge terms reduce to
\be
\mu\exp(i{g^2\over 4\pi T}N\tilde\chi).
\label{muw}
\ee
We then easily get
\be
T_{NC}={g^2N\over 16\pi (N-1)}
\ee
So $T_{NC}$ decreases with $N$. Perhaps surprisingly, we therefore find
that as $N$ becomes larger the two temperatures never cross, and in fact 
the difference between them grows. Nevertheless the temperature
at which the
scaling dimensions of the two operators are equal always stays
equal to the geometrical mean of the two temperatures 
${g^2\over 4\pi}$, in exact agreement with the analysis in the full 
theory eq.(\ref{lowenergy2}).

Why does this happen? 
If we were to allow only the vortices that preserve
the condition $\chi_i=\chi_j$, the only perturbations involving the dual fields
would be
of the form $\mu\exp(i{g^2\over 4\pi T}N(N-1)\tilde\chi)$. This indeed would
lead to much higher $T_{NC}$ so that for $N>4$ the $T_{BKT}$ and $T_{NC}$
would cross. 
However the Lagrangian eq.(\ref{lowenergy2}) contains perturbations which 
create vorticity of a single phase field $\chi_i$, and thus effectively 
violate the equality of all phases. Another way of looking at it is to think 
of the field $\chi$ in eq.(\ref{lred}) as the average field
 $\chi=\sum_{i=1}^{N-1}\chi_i/N-1$. 
The perturbations in eq.({\ref{lowenergy2}) 
then induce fractional vorticity
$2\pi/(N-1)$. The corresponding operators are more relevant than those
with vorticity one and thus the temperature $T_{NC}$ is lower than one would
naively expect. This effect is obviously due to the presence of the $N-1$ 
independent fields all of which can be excited independently.
Thus even though at low temperature the effective theory had only one light 
field, all fields are important in the transition region. 

The preceding discussion is of course only illustrative,
since it neglects the effects of the lightest $W$ bosons.
Those light bosons 
lead to large monopole fugacity $\xi=\exp\{-4\pi M_W/g^2\}$, 
which has an effect of freezing some of the phases 
of the vortex fields. However at finite temperature it is these same $W$ bosons
which are produced more copiously than the others due to their 
relatively large fugacity
$\mu=\exp\{-M_W/T\}$. The appearance of these $W$ bosons however tends to
disorder precisely the same phase fields which are frozen by the 
corresponding 
monopole term by imposing non-vanishing vorticity on them. 
Thus the behaviour of the theory at criticality will be strongly affected
by the presence of these particles and can not be directly deduced
from the effective theory of only one scalar field, even allowing for 
fractional vorticity.

It is interesting to note, that if we go high enough above the critical
temperature where the monopole terms are irrelevant and can be neglected,
the theory is described again quite well in terms of one light field. In this
regime the large fugacity of light $W$'s leads to dynamical constraint
$\tilde\chi_i=\tilde\chi_j$ and we have the theory of one light dual field.

\section{Relating to pure Yang-Mills.}
Although our analysis is not directly relevant to pure Yang Mills theory,
it can be cast in the form which suggests that the relation 
exists and indeed may be closer than apparent at the first glance.

The high energy phase of the Yang-Mills theory is indeed customarily
described in terms of $N-1$ light fields. Those are
the phases associated with the eigenvalues of the Polyakov 
loop, $P$
\cite{ploop}.
Since $P$ is a special unitary matrix, it has $N-1$ independent eigenvalues.
In fact these phases - the components of scalar potential $A_0$, are directly
related to the dual fields $\tilde\eta_i$ of eq.(\ref{lowenergy2}) 
\cite{dunne}. The dual fields $\tilde\eta_i$ appear in the
last term of eq.(\ref{lowenergy2}). This term is nothing
but the free energy of the charged particles $W$. This free energy is usually
expressed in terms of $P$. In the regime where the Higgs expectation value
is large and $W$'s are heavy, the only light components of the vector 
potential are the diagonal ones. Hence in this
regime the Polyakov loop is naturally diagonal. The free energy of a charged
particle with the set of Abelian charges $\vec\alpha$ is then
given by the product of the appropriate eigenvalues of $P$. 
Comparing this with
the last term of eq.(\ref{lowenergy2}) we have
\be
A_0^i={g^2\over 4\pi T}\vec w_i\vec{\tilde\eta}
\ee
where $A_0^i$ is the phase of the $i-th$ eigenvalue of $P$.
Remembering the the following relations between the roots and the weights 
of $SU(N)$
\be
\vec\alpha_{ij}=\vec w_i-\vec w_j, \ \ \ \ \ 
\sum_{i=1}^Nw^a_iw^b_i={1\over 2}\delta^{ab}
\ee
we can rewrite the effective Lagrangian in the hot phase (where the monopole
terms are irrelevant) as
\be
{T\over g^2}\sum_{i=1}^{N}(\partial_\mu A_0^i)^2+
\sum_{ij}\mu_{ij}cos(A^0_i-A^0_j)
\label{ym}
\ee
The phases $\exp i\{A_0^i-A_0^j\}$ are eigenvalues of ${\cal P}$,
where ${\cal P}$ is the Polyakov loop in the adjoint 
representation.
Eq.(\ref{ym}) is therefore of the form similar to the 
``effective action'' discussed in the framework of hot QCD.
Thus the ``effective potential'' in our case 
is given by a linear combination of the eigenvalues of the adjoint 
Polyakov loop.
In fact, at the fixed point where all the fugacities are equal interestingly
enough the potential term generated by $W$ - can be written simply as
\be
\mu{\rm Tr}{\cal P} \ \ .
\ee
In the 
hot Yang Mills theory on the other hand the effective potential is given by
the Bernoulli polynomial \cite{ploop}. The origin of this difference is of
course the large mass of $W$ bosons in the GG model. 
The partition function of a heavy charged particle is well approximated by
the Polyakov loop.
Our derivation 
corresponds to the leading term in the low temperature expansion (expansion
in powers of the Boltzmann factor) which in the GG model
is valid even far above the critical
temperature. In pure Yang Mills on the other hand the ``charged particles''
 - gluons, are massless. As a result the particles are relativistic
and their partition function is not given by the Polyakov loop. Also
the low temperature as such does not exist, and the standard perturbative
calculation corresponds to the genuine high temperature expansion.

Nevertheless it is interesting to observe, that some quantities
calculated in the GG model behave in a way very similar to that in QCD.
In particular consider the ratio of the longest correlation length 
in the sectors with total vorticity $k$, $1/m_k$ to that of vorticity 1.
By total vorticity we mean the quantum number with respect to the magnetic 
$Z_N$ symmetry. This correlation length can be extracted from the
correlation functions of products of $k$ vortex operators 
$<V_{i_1}...V_{i_k}>$. In general this calculation is quite laborious since
the different vortex operators are not degenerate. However they do become
degenerate on the trajectory leading to the fixed point, 
where all the fugacities are equal.
As explained in \cite{KOVNERZN, dunne} at high temperature
the inverse correlation 
length in the vortex channel is given by the ``wall tension'' of the $Z_N$ 
domain wall - solution of the equations of motion for the fields $A_0^i$
with boundary conditions
\be
\exp\{i A_0^i(x)\}\rightarrow_{x\rightarrow -\infty}1, \ \ \ \ 
\exp\{i A_0^i(x)\}\rightarrow_{x\rightarrow \infty}\exp\{i2\pi k/N\}
\label{bound}
\ee
where $x$ is the coordinate transverse to the ``wall''.
In the pure Yang Mills theory the result of this calculation is \cite{pierre}
\be
{m_k\over m_1}={k(N-k)\over N-1}
\label{kwall}
\ee

The equations of motion for the Lagrangian eq.(\ref{ym}) are (we take all 
variables to depend only on one coordinate)
\be
{2T\over g^2}{d^2\over d x^2}(A_0^i-A_0^N)+
\sum_{j\ne i}\mu_{ij}\sin [A_0^i-A_0^j]-
\sum_{j\ne N}\mu_{Nj}\sin [A_0^N-A_0^j]=0
\ee
We are unable to solve these equations in the general case. However in two
special cases they are easy to analyse. Consider first the case discussed in
the previous subsection, when only one of the fields $A_0^i$ is light.
Then obviously on the solution we must have $A_0^i=A_0^j=A$, $i,j,=1,.., N-1$.
Since $A_0^i$ are phases of the eigenvalues of the special unitary
matrix, the last component must then be $A_0^N= (1-N)A$.
These relations must hold for the solution with any $k$, including $k=1$.
Then $A$ satisfies the equation
\be
{2TN\over g^2}{d^2\over d x^2}A+\mu\sin [NA]=0
\ee
with $\mu=\sum_{j\ne N}\mu_{Nj}$.
This is the equation for one scalar field with potential $\cos[NA]$. In 
this case clearly as long
as $k\le N-k$,  the solution for $k\ne 1$ is just the set of $k$ well
separated solutions for $k=1$. When $k\ge N-k$, the same boundary condition
eq.(\ref{bound}) can be satisfied by having $N-k$ walls. 
Thus the tension of the $k$-fold wall is
\be 
m_k=min\{k, N-k\}m_1
\label{kv}
\ee

The other simple case is when all the fugacities are degenerate. Then
following \cite{pierre} we 
can try the following ansatz for solution
\be
A_0^i=kA, \ \ i=1,...,N-k\\
A_0^i=(k-N)A, \ \ i=N-k+1,...,N
\ee
The resulting equation for $A$ is 
\be
{2T\over g^2}{d^2\over d x^2}A+\mu\sin [NA]=0
\ee
This does not depend on $k$. The tension for such a solution
scales as does the kinetic term\cite{pierre} as $k(N-k)$.
Thus the wall tension and the inverse correlation length in the 
channel with vorticity $k$ scales like in hot Yang Mills theory
according to eq.(\ref{kwall}).

Thus even though generically the ratio $m_k/m_1$ in the GG model is not
universal, and depends on the details of the masses of the $W$-bosons,
close to criticality it follows exactly the same simple formula as in 
hot QCD. 

We can analyse in precisely the same way the behaviour of
the ratios of the string tensions of $k$-strings below the transition 
temperature. Due to the self duality of the fixed point, the effective 
Lagrangian in terms of the phases of the vortex operators $\chi_i$ is
identical to the Lagrangian for $A_0^i$ with the substitution $\mu\rightarrow
\zeta$, ${4\pi T\over g^2}\rightarrow {g^2\over4\pi T}$.
The tension of the confining string is then calculated as the tension of
the domain wall separating vacua with different values of 
$\chi_i$ \cite{thooft1}.
We thus find that the ratios of the string tensions also follow the relation
eq.(\ref{kwall}). In fact this scaling relation is commonly known under the
name of ``Casimir scaling'' and is observed to hold for the ratios
of the string tensions in pure Yang-Mills theory at low 
temperature \cite{mike} in both four and three dimensions.

\section{Conclusions}
An interesting feature of our result is that the critical temperature
in the $SU(N)$ theory at large $N$ is proportional to the coupling $g^2$ and 
not to 't Hooft coupling $\lambda=g^2N$. Thus at large $N$ the critical 
temperature approaches zero. The physical reason for this is easy to 
understand. At large $N$ and fixed $\lambda$ the Higgs VEV should also 
scale with $N$ in such a way that the mass of $W$ bosons remains fixed.
The monopole action then grows as $N$ and the photons get progressively
lighter (exponentially with $N$)\footnote{This 
is analogous to the situation in QCD where the instantons become less
relevant at large $N$ and the $\eta^\prime$ meson becomes massless. 
The major difference is of course that while the $\eta^\prime$ mass in QCD
decreases as $1/N$, the photon masses in GG model decrease exponentially.
This difference is due to the non diluteness of the instanton 
gas in QCD as opposed to diluteness of the monopole gas in the GG model.}. 
Thus the thickness of the confining string grows and the density 
of $W$ bosons needed to restore the symmetry becomes smaller and smaller.

More importantly, our main
conclusion is that the deconfining transition in the $SU(N)$ GG model
is second order and the universality class is determined by the infrared fixed
point eq.(\ref{fixedp}). This point is $Z_N$ symmetric and self dual.
We have given some arguments
supporting the possibility that  the fixed point theory is the $Z_N$
parafermionic model \cite{zamolodchikov} although  we were not able
to prove this explicitly. We can however
definitely 
exclude Potts and Villain  universality classes.
In this context we also note that the ratios 
of the ``wall tensions'' calculated in the previous section
(eq.(\ref{kwall})) for $N>3$ are different from 
the corresponding ratios in Villain model (which follow eq.(\ref{kv})
as well as in Potts model (where all the tensions are 
equal $m_k=m_1$)\footnote{Technically speaking
the calculation of the previous section is valid only far enough from 
criticality, so that the monopole terms could be neglected. We believe
however that due to the self-duality of the fixed point the same behaviour 
will also survive in the critical region.}.
This again tentatively 
supports our expectation that the universality class of the 
GG model is different.

To answer this question one should study (numerically or analytically)
the class of $Z_N$ invariant spin systems which has not been studied so far.
The Lagrangian of the relevant model can be taken as eq.(\ref{vlag}).
This is an explicit Lagrangian of $N-1$ interacting phase fields 
which can be easily discretized to define a lattice $Z_N$ invariant 
spin system. Hopefully the $W_N$ symmetry
 of the $SU(N)_1$ WZNW model can be of help here too.

Interestingly, contrary to naive universality arguments the transition is
neither first order as in the 
$N$-state Potts model, nor in the $U(1)$ universality
class as in the Villain model. We believe the reason is precisely the
large number of light fields present in the theory. It is well known that
oftentimes the symmetry alone does not fix the universality class 
of the transition, the number of light fields being the other important 
element.

An interesting question is of course what happens in the pure Yang-Mills 
theory. The global symmetry associated with the phase transition is still
$Z_N$ \cite{KOVNERZN}. The crucial question is what is the number of light 
degrees of freedom. We think there is some grounds to believe that the
description presented in this paper is relevant in this case too.

As discussed in the previous section 
there is direct correspondence in the hot phase between the light 
fields in the GG model and in the pure Yang-Mills theory. 
Again, the usual lore is that the behaviour of these same 
fields $A_0$ at critical temperature determine the universality class of the 
transition. 
Moreover, the ratios of the vortex correlation lengths as well as
string tensions close to criticality in the GG model seems to be similar
to pure Yang Mills theory.
This point of view would then fit with the proposition
that the critical behaviour of the pure Yang-Mills theory is the same as that
of the $SU(N)$ GG model. Of course, universality arguments can never exclude
the possibility of first order transition which can be forced
upon the system by a heavy sector \cite{rob}.
It would be interesting to investigate this question numerically
by lattice gauge theory methods.

\leftline{\bf Acknowledgements}
The work of IIK and BT is supported by PPARC rolling grant 
 PPA/G/O/1998/00567.  
AK is supported by PPARC.                
We are grateful to Chris Korthals Altes for 
encouraging us to study this problem and for very useful discussions on the
behaviour of spin systems and the 't Hooft/Wilson loop ratios. 
We also thank  Andrea Cappelli for discussions about parafermions and 
Mike Teper and Biaggio Lucini for discussions on Casimir
scaling.

\vskip 1cm




\myend
\end{document}